\documentclass{amsart}%PREPRINT

\pdfoutput=1

\usepackage{amsmath,amssymb}

\usepackage{graphicx}

\usepackage{url}

\usepackage{fullpage}

\usepackage[dvipsnames]{xcolor}

\usepackage{hyperref}

\usepackage{algorithm}
\usepackage{algpseudocode}

\usepackage{balance}
\DeclareGraphicsExtensions{.pdf, .jpg,.png,.jpeg}

\usepackage{subcaption}

\newcommand{\Def}{\overset{\text{def}}{=}}
\newcommand{\eps}{\varepsilon}

\title{Congestion Barcodes: Exploring the Topology of Urban Congestion Using 
Persistent Homology}

\author{Yu Wu}
\address{Departments of Physics and Mathematics, 
University of Illinois at Urbana--Champaign, Urbana, IL 61801}
\email{yuwu6@illinois.edu}

\author{Gabriel Shindnes}
\address{Department of Mathematics, University of Illinois at Urbana--Champaign, 
Urbana, IL 61801}
\email{shindne2@illinois.edu}

\author{Vaibhav Karve}
\address{
Department of Mathematics, University of Illinois at Urbana--Champaign,
Urbana, IL 61801}
\email{vkarve2@illinois.edu}

\author{Derrek Yager}
\address{
Department of Mathematics, University of Illinois at Urbana--Champaign,
Urbana, IL 61801}
\email{yager2@illinois.edu}

\author{Daniel B. Work}
\address{Department of Civil and Environmental Engineering and the 
Coordinated Science Laboratory, University of Illinois at Urbana--Champaign, 
Urbana, IL 61801}
\email{dbwork@illinois.edu}

\author{Arnab Chakraborty}
\address{Department of Urban and Regional Planning, 
University of Illinois at Urbana--Champaign, Champaign, IL 61820}
\email{arnab@illinois.edu}

\author{Richard B. Sowers}
\address{Department of Industrial and Enterprise Systems Engineering,
Department of Mathematics, University of Illinois at Urbana--Champaign, 
Urbana, IL 61801}
\email{r-sowers@illinois.edu}

\thanks{\copyright\ 2017 IEEE. Personal use of this material is permitted. 
Permission from IEEE must be obtained for all other uses, in any current or 
future media, including reprinting/republishing this material for advertising 
or promotional purposes, creating new collective works, for resale or 
redistribution to servers or lists, or reuse of any copyrighted component of 
this work in other works. 
To appear in the proceedings of the IEEE ITSC 2017. 
This material is based upon work supported by the Illinois Geometry Laboratory, 
the Program for Interdisciplinary and Industrial Internships at Illinois, and 
the Siebel Energy Institute.}

\begin{document}

\begin{abstract}
This work presents a new method to quantify connectivity in transportation 
networks. 
Inspired by the field of topological data analysis, we propose a 
novel approach to explore the robustness of road network connectivity in the 
presence of congestion on the roadway. 
The robustness of the pattern is 
summarized in a \textit{congestion barcode}, which can be constructed 
directly from traffic datasets commonly used for navigation. 
As an initial demonstration, we illustrate the main technique on a publicly 
available traffic dataset in a neighborhood in New York City.
\end{abstract}

% \maketitle must come after abstract
\maketitle

\section{Introduction}
\subsection{Motivation} 
Not only are today's cities challenged with an excess of congestion, they now 
are faced with an array of routing algorithms, apps, and other tools which can 
easily lead to unexpected and emergent behaviors. 
On the other hand, with more 
data now available, it is becoming possible to think of big-data approaches to 
understanding large-scale mobility problems and compare cities~
\cite{DeriMoura2015,zhu2016using,Zhan2014,guan2016tracking,alonso2017demand,ferreira2013visual}.

Our effort here is to understand the \emph{topology of congestion}.  
Road infrastructure, passenger travel demands, and real-time information 
systems all can interact to create complex behaviors on
road networks. 
An important analytical challenge is to construct coarse 
(e.g., low dimensional) descriptions of the transportation system from the 
high dimensional datasets, from which one can then draw conclusions and make 
data driven decisions. 
Our interest is to apply some recent characterizations 
of general networks to understand large-scale behaviors, with the hope it will 
eventually support new ways to understand and compare mobility behaviors.

% A typical urban road network may consist of regular networks of streets 
% punctuated by high-capacity arteries and irregularly modified by one-way 
% restrictions and complex intersections. 

\subsection{Problem statement and related work}
The focus of the present work is to uncover the topology of traffic congestion. 
This is relevant in addressing questions about its spatial variability. 
For example, trips between OD regions might pass through congested road 
segments, or multiple ``islands'' of smooth flowing traffic might exist in the 
interior of a set of congested links. 
More broadly, the interrelationship 
between
congestion and connectedness can lead to traffic frustration, emergent 
behaviors, and exacerbated gridlock, and we consequently seek methods to 
quantify this structure.

% It is related to the interest in macroscopic fundamental 
% diagrams~\cite{geroliminis2008existence}, which is an aggregation approach to 
% summarize traffic flow in an urban region. %Other papers on urban traffic 
% estimation include~
% \cite{castro2012urban,hunter2009path,zhan2013urban,santi2014quantifying,zhan2015bayesian}.

Our interest here is applying topological data analysis to traffic behavior and 
predictability.  
To facilitate new approaches to address urban traffic, cities 
(viz. Chicago \cite{Chicago} and New York City \cite{NYC}) are releasing large 
mobility datasets. 
% We would like to develop some notions of persistent homology 
% to provide insight to traffic networks.  
% Large-scale topological analyses seem 
% like a natural tool to use in understanding traffic networks. 
We believe that this line of inquiry may give some new types of useful insights 
which ``coarse-grain" behaviors in such a way that we can seek to compare cities 
and transfer quantitative knowledge from one place to another. 
%(see also \cite{ahmed2014local}).

A variety of topological information can be studied.  
The simplest ``homology" counts the number of connected
components in the topology.  
One can also construct homologies reflecting the structure of \emph{paths}; 
this has implications for robotic path planning in the face of uncertainty 
\cite{bhattacharya2015persistent}
and also connectivity in the brain \cite{lee2012persistent}.
By way of precedence, the article \cite{ahmed2014local}, uses persistent 
homology to measure similarities between road maps.

Persistent homology, introduced by \cite{edelsbrunner2000topological} and 
\cite{MR2121296}, (see also \cite{MR2121296, Taylor2015, 
bhattacharya2015persistent, edelsbrunner2008persistent}) seeks to ``unfold" 
geometries of datasets, giving new insights into the structure of the data and 
the robustness of this structure.  
Many datasets naturally have an unfolding parameter, and one can compute
how various topological objects (characterized as null spaces, ranges, and other 
linear-algebraic objects) appear and disappear as the parameter changes.  
Topological objects which persist over large ranges of parameter values can 
be interpreted as stable and robust, while those which are short-lived in 
parameter space might be thought of as noise.  
A \textit{barcode}, described in detail later, allows one to 
graphically capture dependence on parameters.

When applied to traffic networks, these techniques can help us analyze large 
travel datasets and generate useful insights. 
In particular, it provides a way 
to capture the relationships between the following aspects of a transportation 
network with respect to a given speed threshold: number of connected links that 
meet such threshold, sizes of such connected links, and coverage of the region 
given a speed thresholds. 

The original motivation for persistent homology was
in understanding stable aspects of point clouds of data
(see \cite{ghrist2008barcodes}).  
In that case, a distance parameter $\eps$ can be used
to connect points (the \v Cech or Rips complexes).  
As the parameter $\eps$ is increased, more points are included in the simplices; 
understanding the robust aspect of these simplices allows one to
reconstruct the essential features of the point cloud
in ways which are somewhat impervious to noise.
This has been adapted to image processing \cite{carlsson2008local} and analysis 
of coverage of sensor networks \cite{dlotko2012distributed}, and the structure 
of some data in global development \cite{banman2017mind}.  
Persistence of more complicated geometric invariants have been used to study 
feasibility sets in robotics; see~\cite{MR2358377}.

\subsection{Outline and contributions}
The main contribution of this work is the application of persistent homology to 
understanding road traffic networks. 
To our knowledge, this is the first 
application of topological data analysis to traffic.  
Our hope is that this line of research will lead to some new and robust 
techniques of capturing the structure of congestion in ways which can be used 
to compare different cities and regions.  

In Section ~\ref{sec:background}, we provide a brief introduction of persistent 
homology and explain how it is applied to road networks. 
In Section~\ref{sec:algorithm}, we provide the algorithms used to construct a 
"congestion barcode", before applying it in  Section~\ref{sec:data} to a small 
example from New York City.

\section{Background on Persistent Homology}\label{sec:background}
We briefly review the main ideas of persistent homology and explain how it can 
be applied to road networks. 
The interested reader is referred to 
\cite{edelsbrunner2008persistent,ghrist2008barcodes} for a detailed description 
and survey of applications.
We develop the notion of persistence for the \emph{simplest} invariant; the 
number of connected components (corresponding to the zero-th Betti number; see 
\cite{munkres1984elements}).  
We will here develop the notion of connectedness 
as, informally, a set of good (e.g., fast) roads, all of which one can traverse 
without ever needing to drive on a bad (slow) road.

Abstractly, we have a weighted undirected finite graph $G=(V,E)$ where $V$ is a 
finite set of vertices, which
we interpret as intersections, and $E$ is the set
of edges, which we interpret as roads or links.
Each edge $e$ is a set $\{v_1,v_2\}$ of two distinct vertices; $e$ connects 
$v_1$ and $v_2$.
Each edge $e$ has a nonnegative weight $W_e$. 
For specificity, we will think of the weights as speeds, so
higher weights correspond to better traffic conditions.  
For simplicity, we assume that each $v\in V$ is in one of the edges 
$e\in E$ (i.e., $G$ has no disconnected vertices).

If $G'=(V',E')$ is a finite graph, we say that a subset $C\subset V'$ is 
connected if, for every pair $v'_a$ and $v'_b$ of vertices in $C$, 
there is a sequence $\{v'_n\}_{n=1}^N$ of vertices
with $v'_1=v'_a$ and $v'_N=v'_b$, such that $\{v'_n,v'_{n+1}\}\in E'$; i.e., 
there is a path leading from $v'_a$ to $v'_b$ along the edges in $E'$.  
We can decompose $V'$ into a finite collection of maximal connected components.

We finally introduce a concept of a \textit{level}, or \emph{persistence 
parameter}, which is used to construct and compare subgraphs. 
For each level $\lambda\in [0,\infty)$, define a subgraph $G(\lambda)$ generated
by edges whose weight is greater than $\lambda$ as
\begin{equation}\label{E:VDef} G(\lambda) \Def  (V_\lambda,E_\lambda), 
\end{equation}
where
\begin{equation} E_\lambda = \{e\in E: W_e> \lambda\}, 
\label{E:subEDef}\end{equation}
and where
\begin{equation} V_\lambda = \bigcup_{e\in E_\lambda}e. 
\label{E:subVDef}\end{equation}

%%%%%%%%%%%%%%%%%%%%%%%%%%%%%%%%%%%%%%%%%%%%%%%%%%%%%%%%%%%%%%%%%%%%%%%%%%%%%%%
%BEGIN: Subsection: Barcodes expained%%%%%%%%%%%%%%%%%%%%%%%%%%%%%%%%%%%%%%%%%%
%%%%%%%%%%%%%%%%%%%%%%%%%%%%%%%%%%%%%%%%%%%%%%%%%%%%%%%%%%%%%%%%%%%%%%%%%%%%%%%
\subsection{Barcodes explained}\label{SS:BC}

The focus of persistent homology is understanding how the topology of 
$G(\lambda)$ changes as $\lambda$ changes.  
To start, we note that 
\begin{equation*} 
G(\lambda_1) \subset G(\lambda_2) 
\end{equation*}
if $\lambda_1\ge \lambda_2$ (i.e., $V_{\lambda_1}\subset V_{\lambda_2}$
and $E_{\lambda_1}\subset E_{\lambda_2}$).  
Intuitively, as $\lambda$ decreases, we include
more links in the definition of $E_\lambda$ (and thus $V_\lambda$).  
The map $\lambda\mapsto G(\lambda)$ is a reverse filtration (the $G(\lambda)$'s 
get larger as $\lambda$ decrease, as opposed to getting larger as $\lambda$ 
increases).

Our goal here is to understand how the connected
components of $G(\lambda)$ change as $\lambda$ decreases and $G(\lambda)$ fills 
in more and more of $G$. 
Informally, we think of the parameter $\lambda$ as 
reversed time (we want to use reversed time, since connected components appear 
as the ``time'' parameter $\lambda$ decreases).  
For $\lambda_1$ and $\lambda_2$ with $\lambda_1>\lambda_2$, we want to 
compare the connected components of $G(\lambda_1)$ with those of $G(\lambda_2)$.  Several things can happen.
$G(\lambda_2)$ can contain a new connected component (a ``birth"). 
Connected components can merge and some of them will ``die". 
Finally, a connected component can grow in size without merging with other 
connected components.

A \textit{barcode} focuses on the evolution the connected components of 
$G(\lambda)$ by mapping each connected component into a different line, or 
\textit{bar}.  
A bar starts when the corresponding component is born, and ends when it merges
(the bar reflects the growth of a connected component as long as it does not 
merge).  
The structure of the different bars will help us visualize how 
connected components of the good parts of the traffic network appear and merge.

We can think of each bar in the barcode as a key-value dictionary which evolves 
with the parameter $\lambda$.  
The bar consists of a \texttt{start} value when 
the barcode was born, an \texttt{end} value when
the bar merged and died (if it has in fact already merged), and a \texttt{state} 
consisting of the current connected component.  
The \texttt{state} of the bar keeps the 
information we need to compare the current (i.e., the current value of 
$\lambda$) connected set to a connected set at a later time.
The barcode is a list of bars.

Focusing on the death of bars, we need to agree on which bar to extend when two 
connected components merge.  
When bars merge, we say that the oldest bar--i.e., 
the one with the highest \texttt{start} value (recall that we are reversing 
time, so ``older" means larger starting values)--survives, while the younger 
ones (the ones with lower starting value) die.

We define the \emph{persistence} of each bar as the absolute difference between 
its \texttt{start} and \texttt{end} $\lambda$-values. 
We represent these bars in a plot with the length of each bar being proportional 
to its persistence, ordering the bars by total length with the longest bars on 
the bottom. we call this plot the barcode diagram, or simply the barcode.

If a component exists for wide range of levels $\lambda$ i.e. its bar has a high 
persistence, we can think of it as robust;
if a component exists only for a short range of parameter
values, one might interpret it as result of noise.

\section{Algorithm}\label{sec:algorithm}
The first step is to decompose each $G(\lambda)$ into connected components.  
A \emph{depth-first search} ~\cite{even2011graph} can efficiently do this; 
see Algorithm \ref{alg:dfs}.
%To fix the calculations, assume that $V=\{v_n\}_{n=1}^N$.
%The depth-first search algorithm \cite{even2011graph} sequences through the 
%vertices of the graph, fixing a vertex
%and finding other vertices connected to it, and %recursively repeating the 
%procedure for these new vertices.  To implement
%this\Dan{what does ``this refer to''? DFS, or the algorithm that returns the 
%set of distinct components in $G(\lambda)$?}, let's construct an adjacency 
%matrix 
%corresponding to $G(\lambda)$ of \eqref{E:VDef} by defining
%\begin{equation*} A_{n,n'}(\lambda) = \begin{cases} 1 &\text{if 
%$W_{\{v_n,v_{n'}\}}>\lambda$} \\
%0 &\text{else;}\end{cases}\end{equation*}
%(i.e., $A_{n,n'}(\lambda)=1$ if and only if $\{v_n,v_{n'}\}\in E_\lambda$);
%the matrix $A(\lambda)$ is symmetric. 
%We can then rewrite \eqref{E:subEDef} and \eqref{E:subVDef} as 
%\begin{align*} V_\lambda &= \lb v_n: \text{the $n$-th row of $A(\lambda)$ is 
%nonzero}\rb \\
%E_\lambda &= \lb \{v_n,v_{n'}\}: A_{n,n'}(\lambda)=1\rb \end{align*}
\begin{algorithm*}
\caption{Create list of connected components of $G(\lambda)$ using depth first 
search}\label{alg:dfs}
\begin{algorithmic}[1]
\State{Label all vertices in $V_\lambda$ as ``unvisited"}\Comment{We relabel the 
vertices as they appear in connected subsets of $G(\lambda)$}
\State{Let $\mathcal{W}$ be a list of connected components of 
$G(\lambda)$}\Comment{Initially empty}

\For {each vertex $v\in V_\lambda$ which is still ``unvisited"}
    \State{Start a new connected component $C$ which contains (is ``rooted" at) 
$v$} \Comment{$C=\{v\}$}
    
    \Function{\texttt{DFS}}{node v'} \Comment{Define recursive function which 
adds new vertices to $C$ and ``visits" vertices}
\For {each unvisited vertex $v^{\prime \prime}\in V_\lambda$ with 
$W_{v',v^{\prime \prime}}>\lambda$} \Comment{$v^{\prime \prime}$ is connected to 
$v'$ in $G(\lambda)$}
    \State{label $v^{\prime \prime}$ as visited}
    \State{add $v^{\prime \prime}$ to $C$}
    \State{\texttt{DFS}($v^{\prime \prime}$)}\Comment{Recursive step}
    \EndFor
    \EndFunction \Comment{\texttt{DFS} has modified $C$ and "visited" vertices}
\State{ \texttt{DFS}(v)}\Comment{Apply \texttt{DFS} to $v$}
\State{Add $C$ to $\mathcal{W}$} \Comment{$C$ is the unique connected component 
of $G(\lambda)$ which contains $v$}
\EndFor
\end{algorithmic}
\end{algorithm*}
\noindent This gives us the connected components of $G(\lambda)$.  We visit each 
vertex in $V(\lambda)$ and recursively find the other vertices which are 
connected by a weight of more than $\lambda$.

One of the computational challenges of constructing barcodes is efficiently 
carrying out the comparisons between the components of the current $G(\lambda)$, 
which we compute via the depth-first search algorithm, and the \texttt{state} 
values (i.e., connected sets) of the bars at the prior values.  
Note that the $G(\lambda)$'s (and thus their connected components) only change 
at the different values of 
  \begin{equation} 
    \Lambda\Def \{W_e: e\in E\}. \label{E:Lvals} 
  \end{equation}
To facilitate merging, let's organize the list of bars in the barcode by start 
time, with larger \texttt{start} time (i.e., older age) being first.  
The merging order outlined in Subsection \ref{SS:BC} thus corresponds to the 
bars with the higher list index merging into the one with the lower list index.  
We carry this out in Algorithm \ref{alg:barcode}; when a bar b dies, we cease 
updating its \texttt{end} value.  
To aid in the clarity of the merging process, we could also, in Algorithm 
\ref{alg:barcode} set the \texttt{state} of a dead bar to $\emptyset$.

\begin{algorithm*}
\caption{Create barcodes.  A bar b is a key-value dictionary consisting of a 
\texttt{start}, \texttt{end}, and \texttt{state}.}\label{alg:barcode}
\begin{algorithmic}[1]
\State{Create ordered list $B$ of bars}\Comment{Initially empty.}
\For {each $\lambda\in \Lambda$ of \eqref{E:Lvals}, ordered by decreasing 
value}\Comment{Reverse time}
  \For{each connected component $C$ of $G(\lambda)$}\Comment{Use \texttt{DFS}}
    \For{each barcode $b\in B$, with oldest bars first}
        \If{$C$ and b[\texttt{state}] intersect}
            \If{we have not yet declared a merge}
                \State{Declare a merge}\Comment{Occurs for oldest intersecting 
bar; extend this bar}
                \State \text{b[\texttt{end}]$\leftarrow \lambda$}\Comment{extend 
bar to at least current time}
                \State \text{b[\texttt{state}]$\leftarrow C$}
            \Else\Comment{$C$ intersects with younger bar}
                \State{bar dies}
            \EndIf
        \EndIf
    \EndFor
    \If{a merge has not yet been declared}\Comment{$C$ has not intersected with 
any bar}
        \State start a new bar with \texttt{start} set to $\lambda$ and 
\texttt{state} set to $C$.
    \EndIf
\EndFor
\EndFor
\end{algorithmic}
\end{algorithm*}
Each connected component can merge several bars, extending only one of them (the 
oldest).  
Since $\lambda \mapsto G(\lambda)$ is a reverse filtration (i.e., 
$\lambda \mapsto G(\lambda)$ is decreasing in $\lambda$), different connected 
components of $G(\lambda)$ cannot affect the same bar.

\section{Description of the dataset: New York City traffic speeds}
\label{sec:data}
\begin{figure}%[hbtp]
\centering
\includegraphics[width=0.5\textwidth]{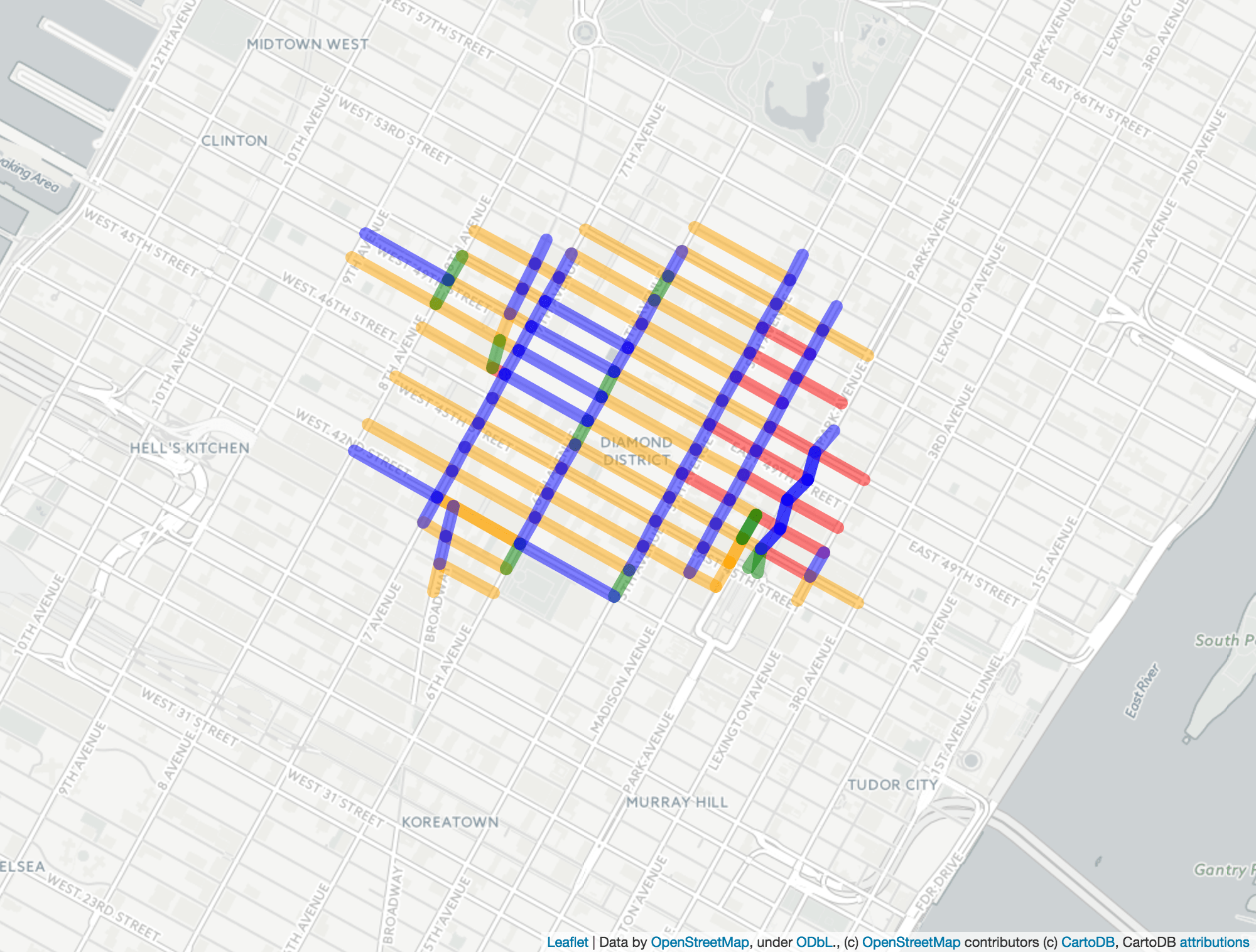}
\caption{Case study network. Diamond District streets (144 links, and 74 nodes) 
in New York City. 
The red roads correspond to speeds in $[0,2)$, the orange 
roads correspond to speeds in $[2,3)$,  the blue roads correspond to speeds in 
$[3,4)$, and the green roads correspond to speeds in $[4,\infty)$.} 
\label{fig:DD}
\end{figure}

We are interested in developing our ideas for a restricted dataset.  
We start with a dataset of 2011 taxi data \cite{NYC} as processed by 
\cite{donovan2015using} and look at traffic speeds in the Diamond district 
between 9 and 10 AM on workdays during June, July and August of 2011. 
Roughly, this dataset should reflect traffic behavior which is fairly 
homogeneous in time, allowing us to focus on spatial fluctuations.  
The roads are highlighted in Figure \ref{fig:DD}. 
We consider $152$ one-directional links, corresponding to 144 roads; 
there are 8 two-way roads.  
The dataset has $D\Def 66$ days in it.

If we let $s_{\ell,t}$ correspond to the speed in link $\ell$ on day $t$, let's 
define
\begin{equation}\label{E:WDef} W_\ell\Def \frac{1}{D}\sum_{d=1}^D s_{\ell,t} 
\end{equation}
as the average speed on link $\ell$.  
To resolve the ambiguity of this definition on the $8$ two-way roads, 
we replace \eqref{E:WDef} by the average 
over both directions during this time; we average over the $132$ speeds 
corresponding to $66$ speeds in one direction, and $66$ speeds in the other.  
This affects only 8 roads, and reflects a somewhat justifiable assumption that 
congestion in one direction may cause congestion, both by proximity and by 
left turns.

\begin{figure}
\centering
\includegraphics[width=0.5\textwidth]{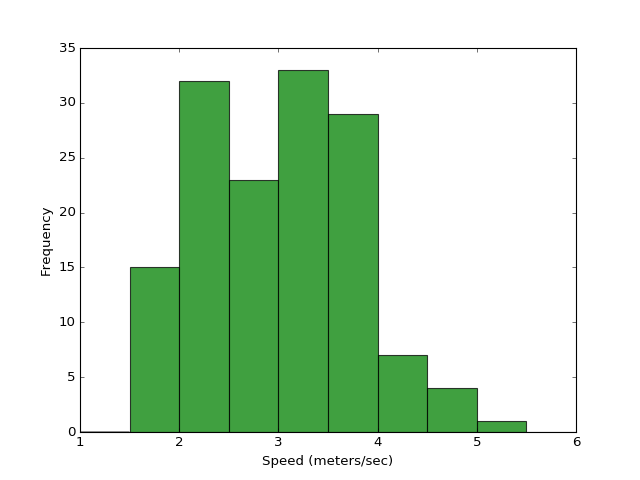}
\caption{Histogram of the average road speed in the Diamond District from 9-10 
AM in the summer of 2011.}\label{fig:hist}
\end{figure}

The $S_\ell$'s, the average speeds on the different links, range from 1.5 m/s to 
5.00 m/s (all speeds are in meters/second), with average of
3.0 m/s and standard deviation of 0.78 m/s. 
The low speeds come from the fact that the area is highly congested in the 
late morning hour from which the data was obtained. 
See Figure \ref{fig:hist} for a histogram of the speed 
distribution on the graph.  

\begin{figure}
\centering
\begin{subfigure}{0.4\columnwidth}
\includegraphics[width=\columnwidth]{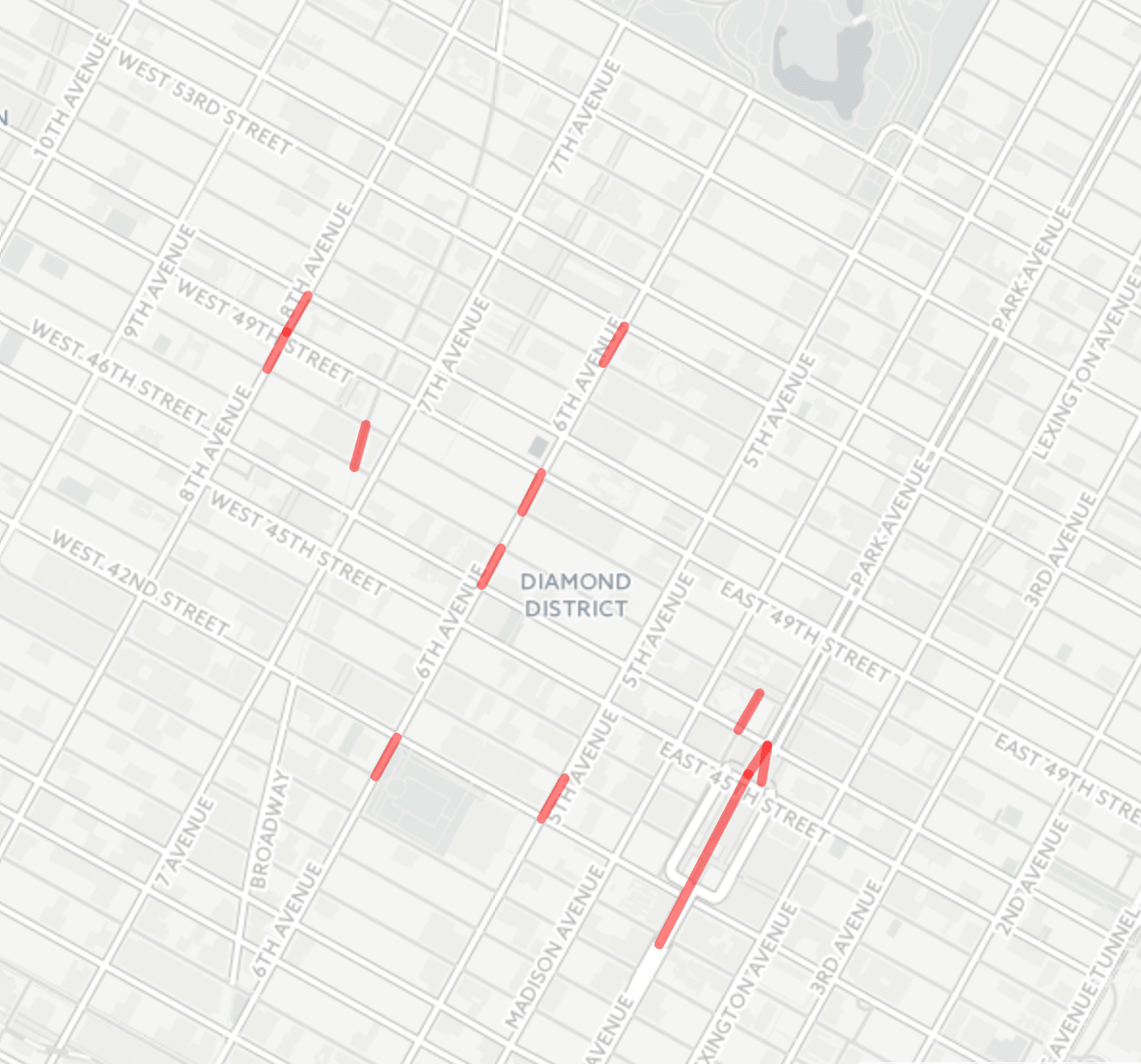}
\caption{Roads with speeds in $[4,\infty)$.}\label{fig:new_4toinf}
\end{subfigure}
\begin{subfigure}{0.4\columnwidth}
\includegraphics[width=\columnwidth]{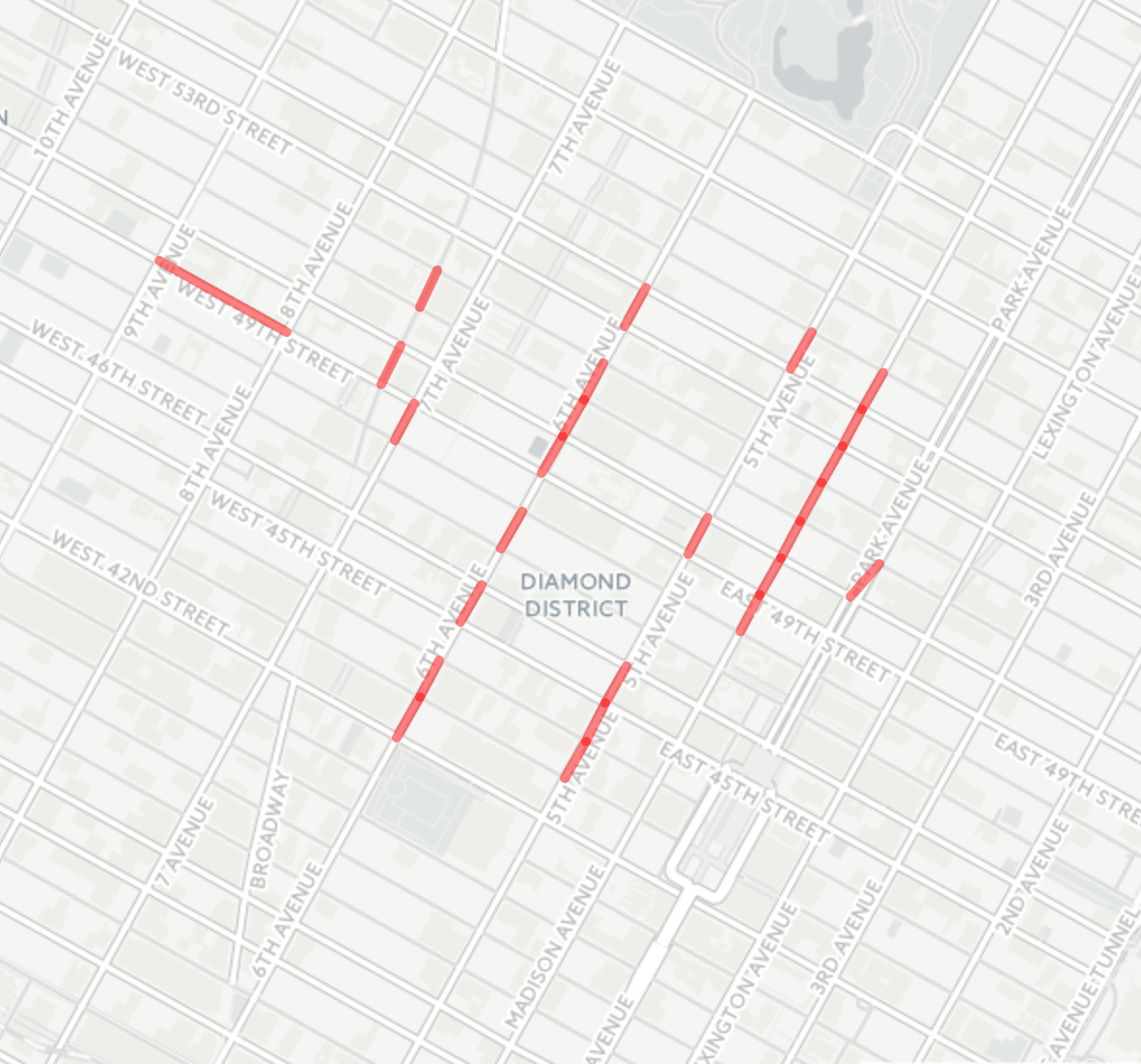}
\caption{Roads with speeds in $[3.5,4)$.}\label{fig:new_3p5to4}
\end{subfigure}
\begin{subfigure}{0.4\columnwidth}
\includegraphics[width=\columnwidth]{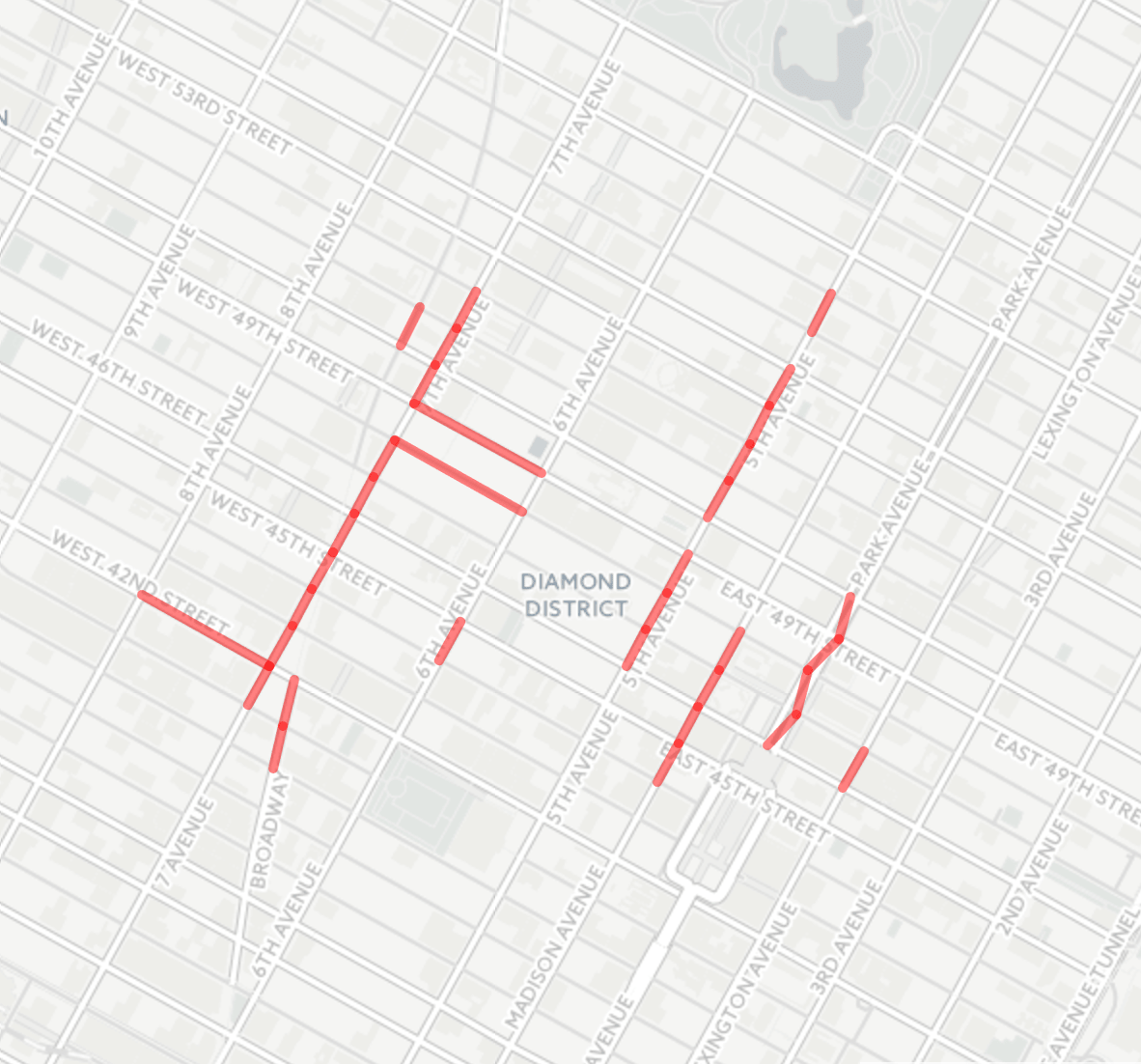}
\caption{Roads with speeds in $[3,3.5)$.}\label{fig:new_3to3p5}
\end{subfigure}
\begin{subfigure}{0.4\columnwidth}
\includegraphics[width=\columnwidth]{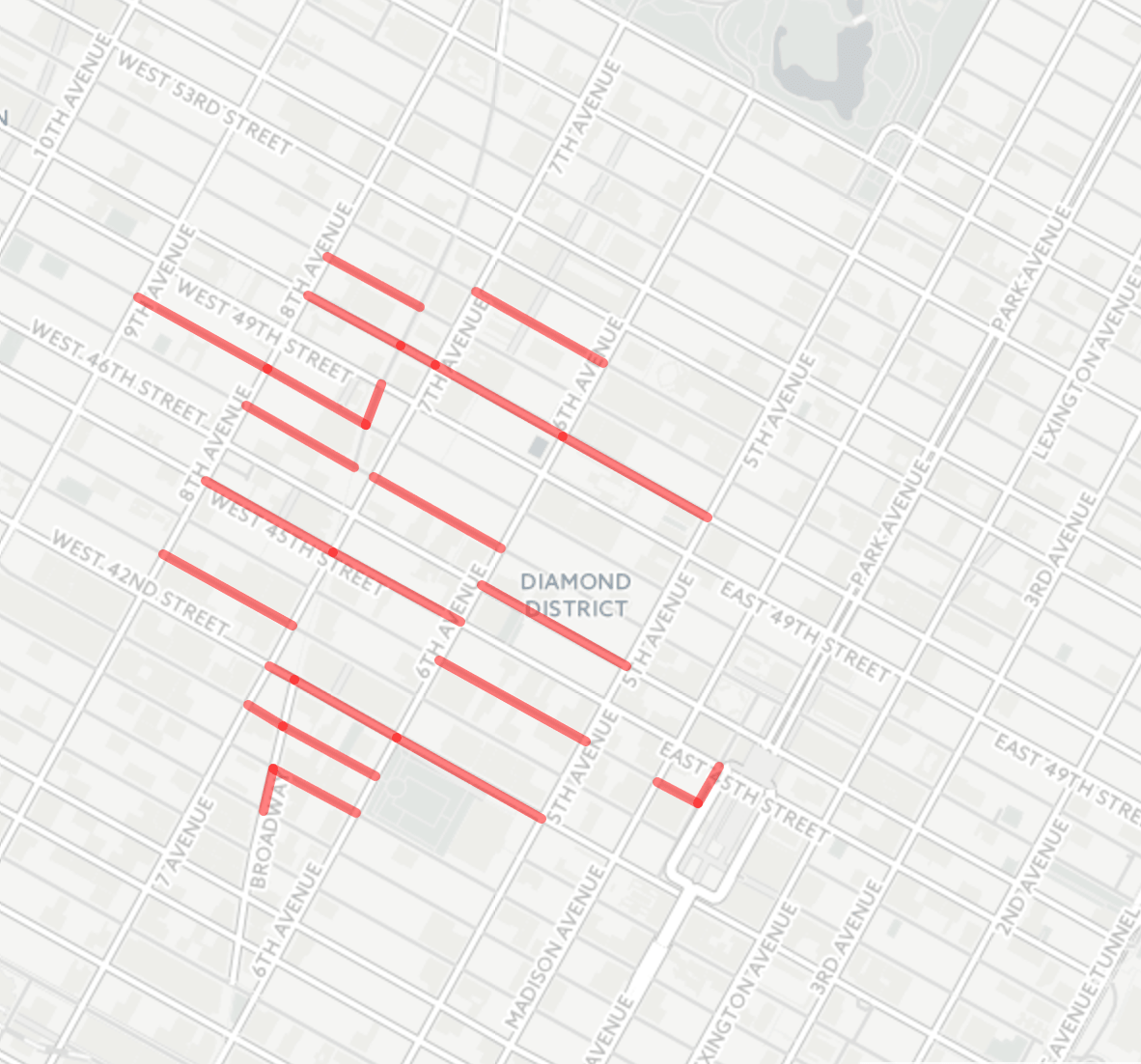}
\caption{Roads with speeds in $[2.5,3)$.}\label{fig:new_2p5to3}
\end{subfigure}
\begin{subfigure}{0.4\columnwidth}
\includegraphics[width=\columnwidth]{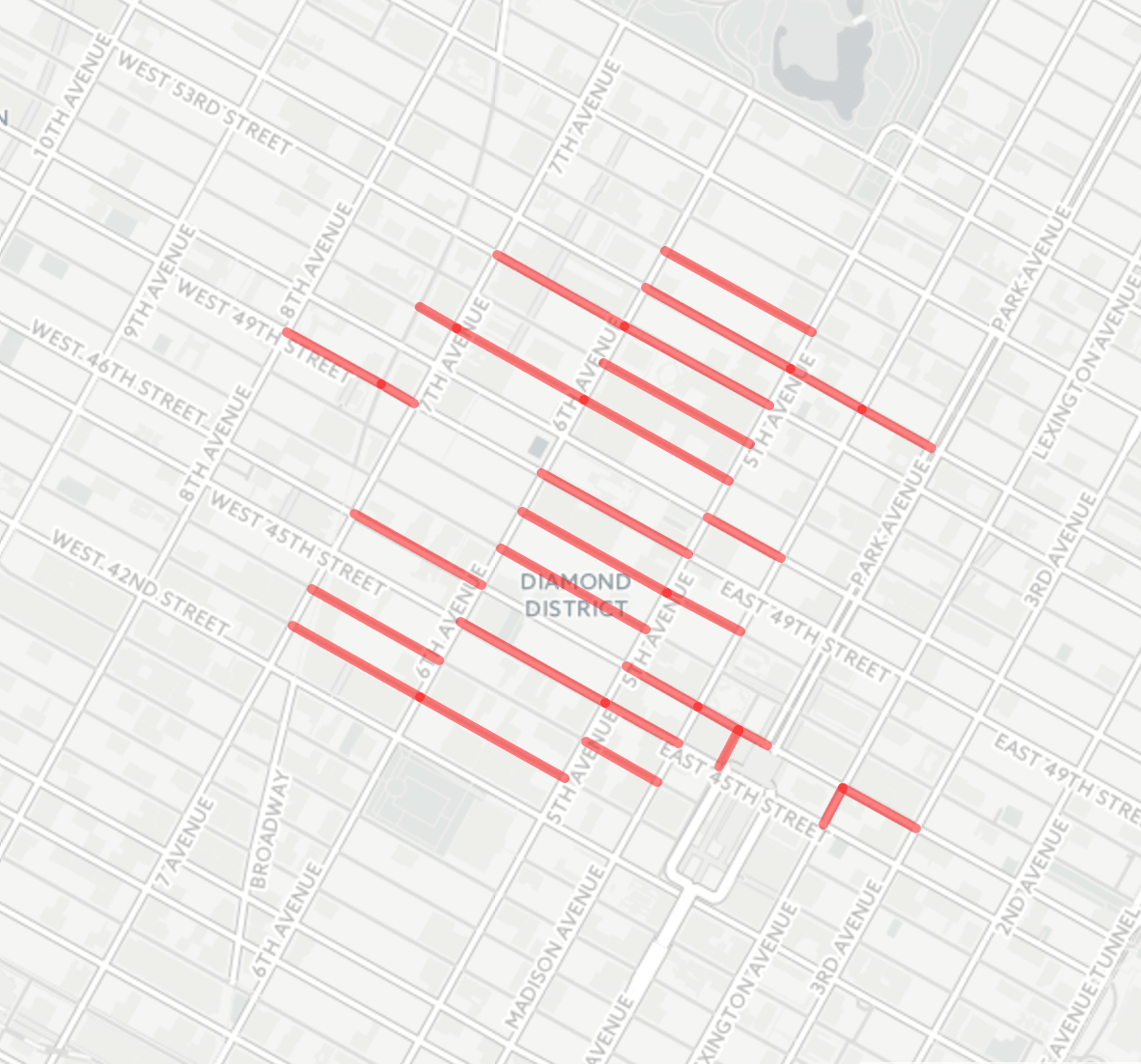}
\caption{Roads with speeds in $[2,2.5)$}\label{fig:new_2to2p5}
\end{subfigure}
\begin{subfigure}{0.4\columnwidth}
\includegraphics[width=\columnwidth]{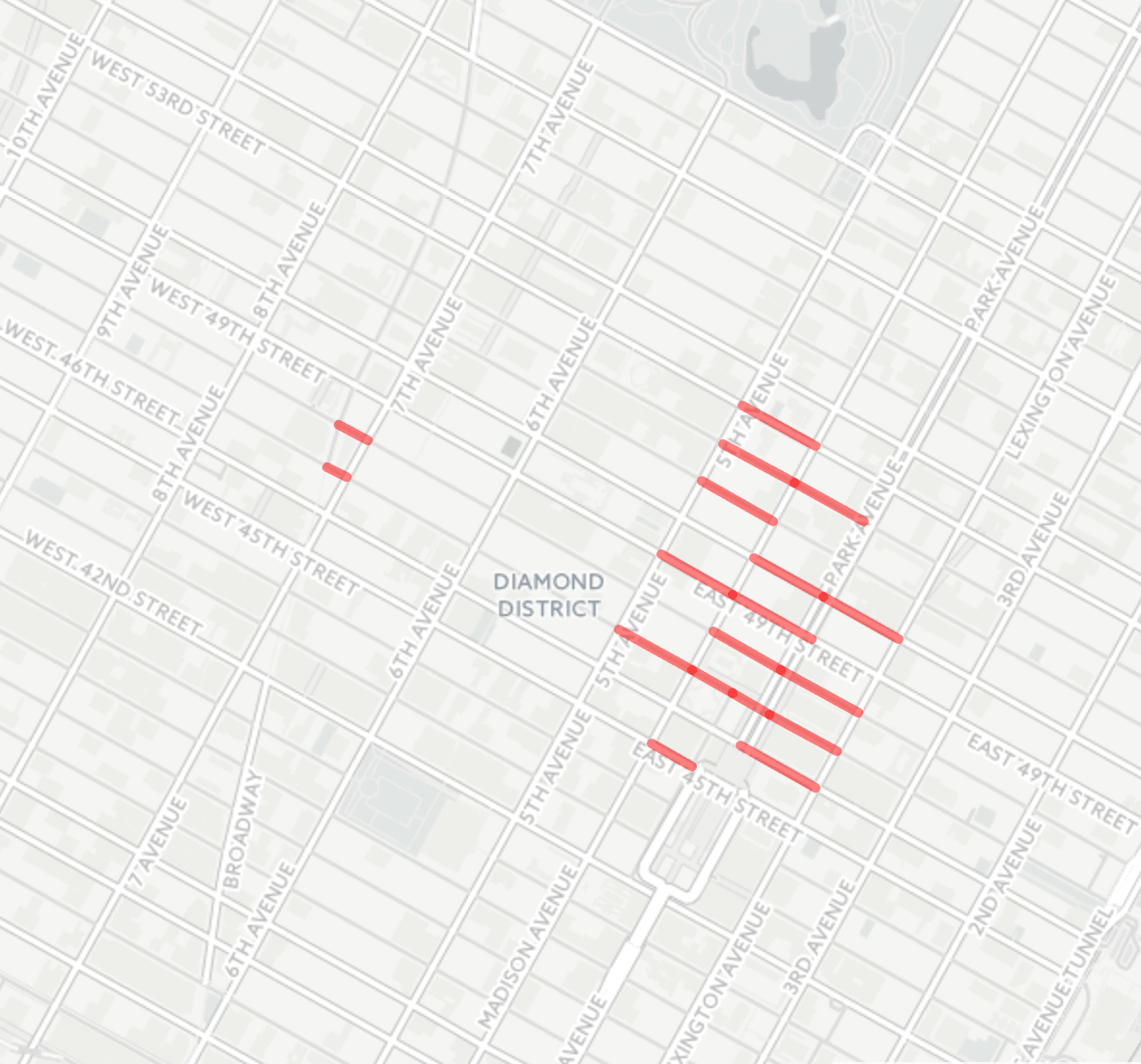}
\caption{Roads with speeds in $(0,2)$}\label{fig:new_0to2}
\end{subfigure}
\caption{Spatial distribution of the average road speed in the Diamond District 
from 9-10 AM in the summer of 2011.}
\end{figure}

We are primarily interested in how these speeds are spatially distributed.  
In Figure \ref{fig:new_4toinf}, the streets with speeds greater than 4.0 m/s are 
highlighted in red; these are the fastest streets. 
In Figure \ref{fig:new_3p5to4}, the streets with speeds between 3.5 and 4 m/s 
are highlighted; these are slightly slower.  
Figure \ref{fig:new_0to2} gives roads with speeds less than 2, while 
Figure \ref{fig:new_2to2p5} gives roads with speeds between 2 and 2.5; 
these are the slowest and second-slowest roads.

Let's look at the connectedness of the roads in Figures 
\ref{fig:new_4toinf}--\ref{fig:new_0to2}.  
Since connectedness of the partially uncongested roads is our primary interest, 
the actual length of the various links plays no role.  
Figure \ref{fig:new_4toinf} has 9 connected components, 
Figure \ref{fig:new_3p5to4} has 14 connected components, Figure 
\ref{fig:new_3to3p5} has 11 connected components and Figure \ref{fig:new_2p5to3} 
has 14 connected components.  
The connectivity of the components gives an idea 
of how "big" a region of good or bad traffic is.  
Informally, traffic congestion refers to a large connected cluster of slow 
roads.  
Most of the NE-SW roads are 
fast (Figures \ref{fig:new_4toinf}--\ref{fig:new_3to3p5}), and the NW-SE roads 
are slower (Figures \ref{fig:new_2p5to3}--\ref{fig:new_0to2}).  
The system seems to become a solid connected component at around 2.5.

\begin{figure}[hbtp]
\centering
\begin{subfigure}{0.4\columnwidth}
\includegraphics[width=\columnwidth]{new_4toinf}
\caption{Roads with speeds in $[4,\infty)$. 
There are 9 connected 
components.}\label{fig:G_4}
\end{subfigure}
\begin{subfigure}{0.4\columnwidth}
\includegraphics[width=\columnwidth]{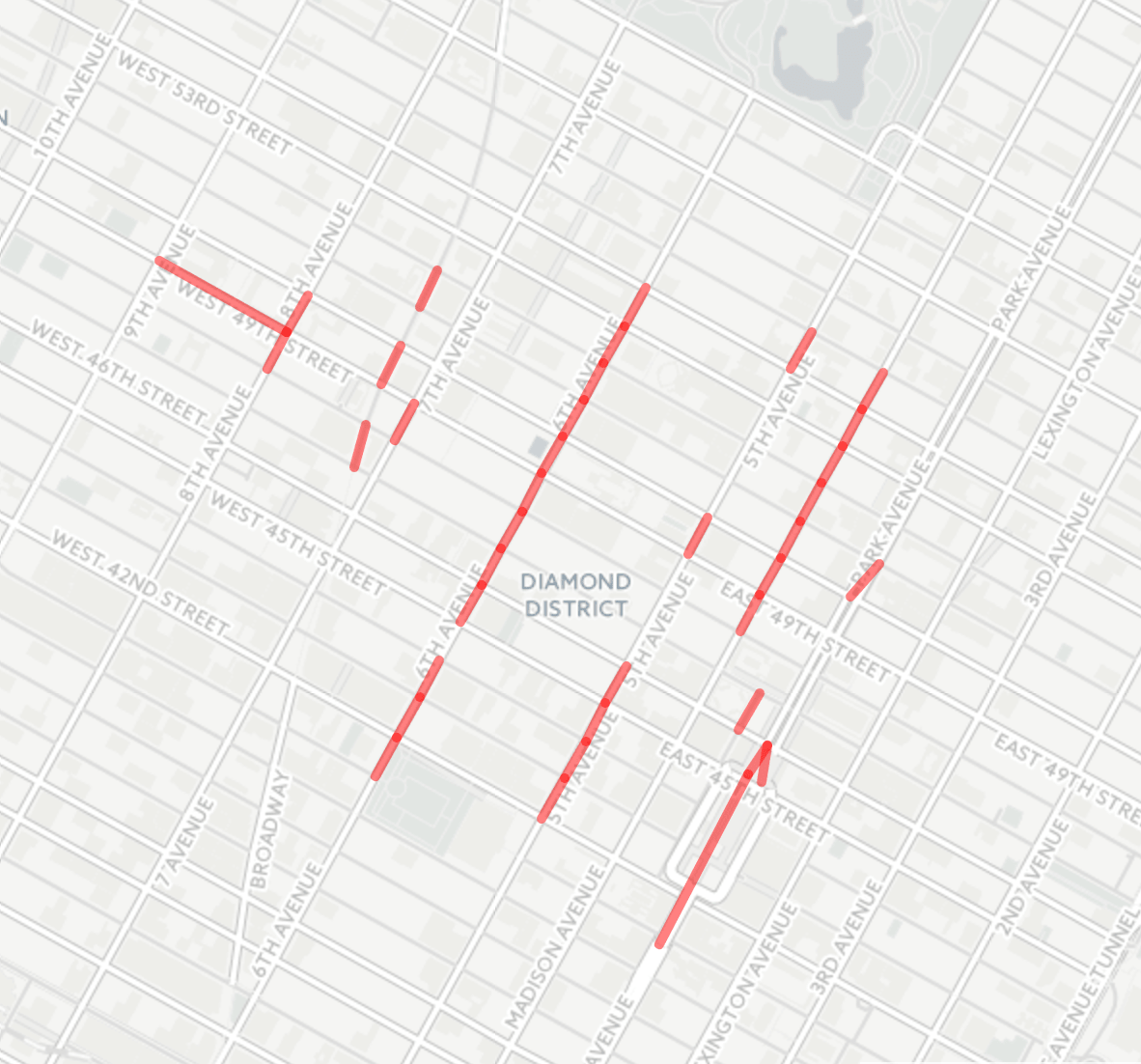}
\caption{Roads with speeds in $[3.5,\infty)$.   There are 14 connected 
components.}\label{fig:G_3p5}
\end{subfigure}
\begin{subfigure}{0.4\columnwidth}
\includegraphics[width=\columnwidth]{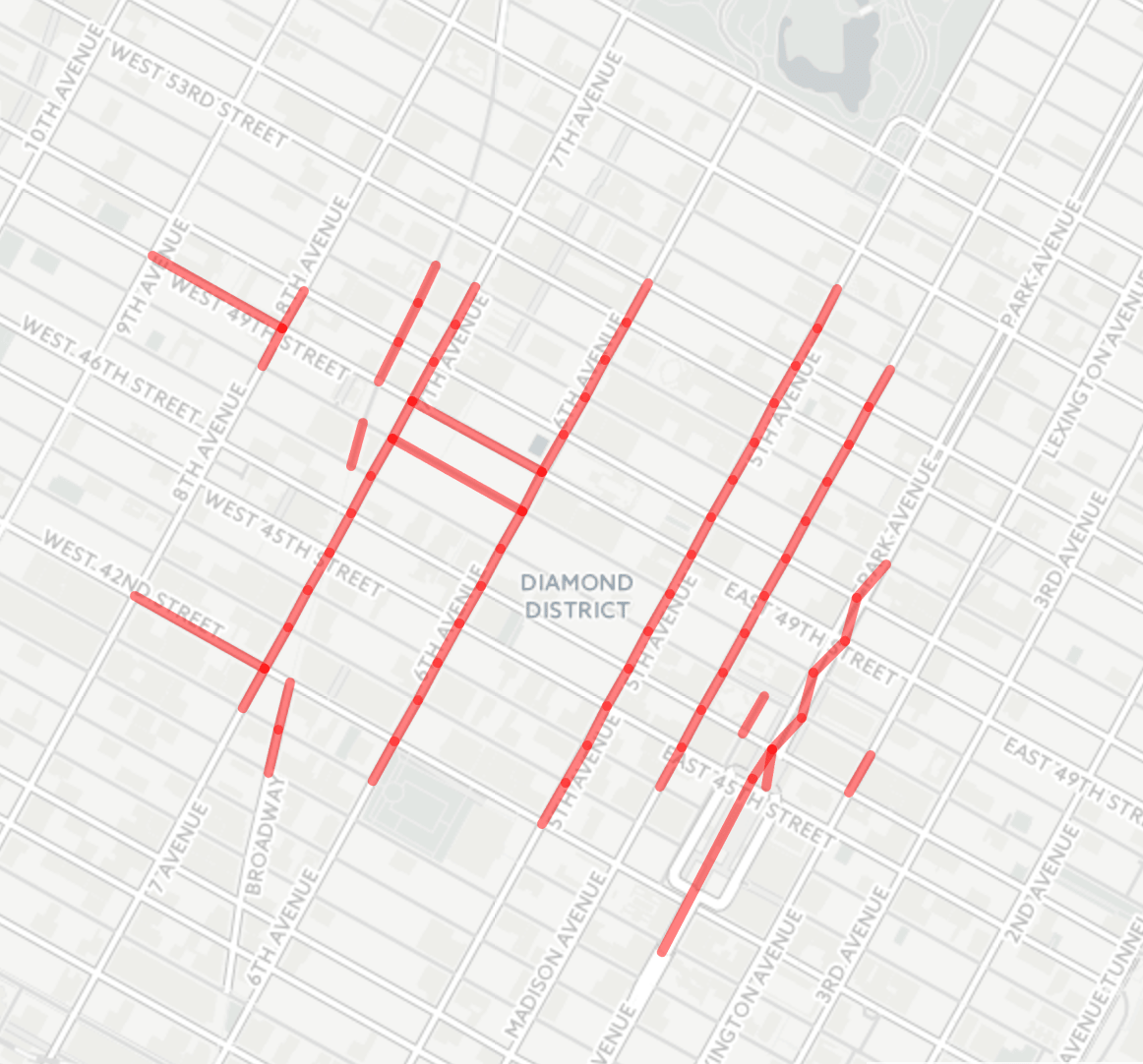}
\caption{Roads with speeds in $[3,\infty)$.  There are 10 connected 
components.}\label{fig:G_3}
\end{subfigure}
\begin{subfigure}{0.4\columnwidth}
\includegraphics[width=\columnwidth]{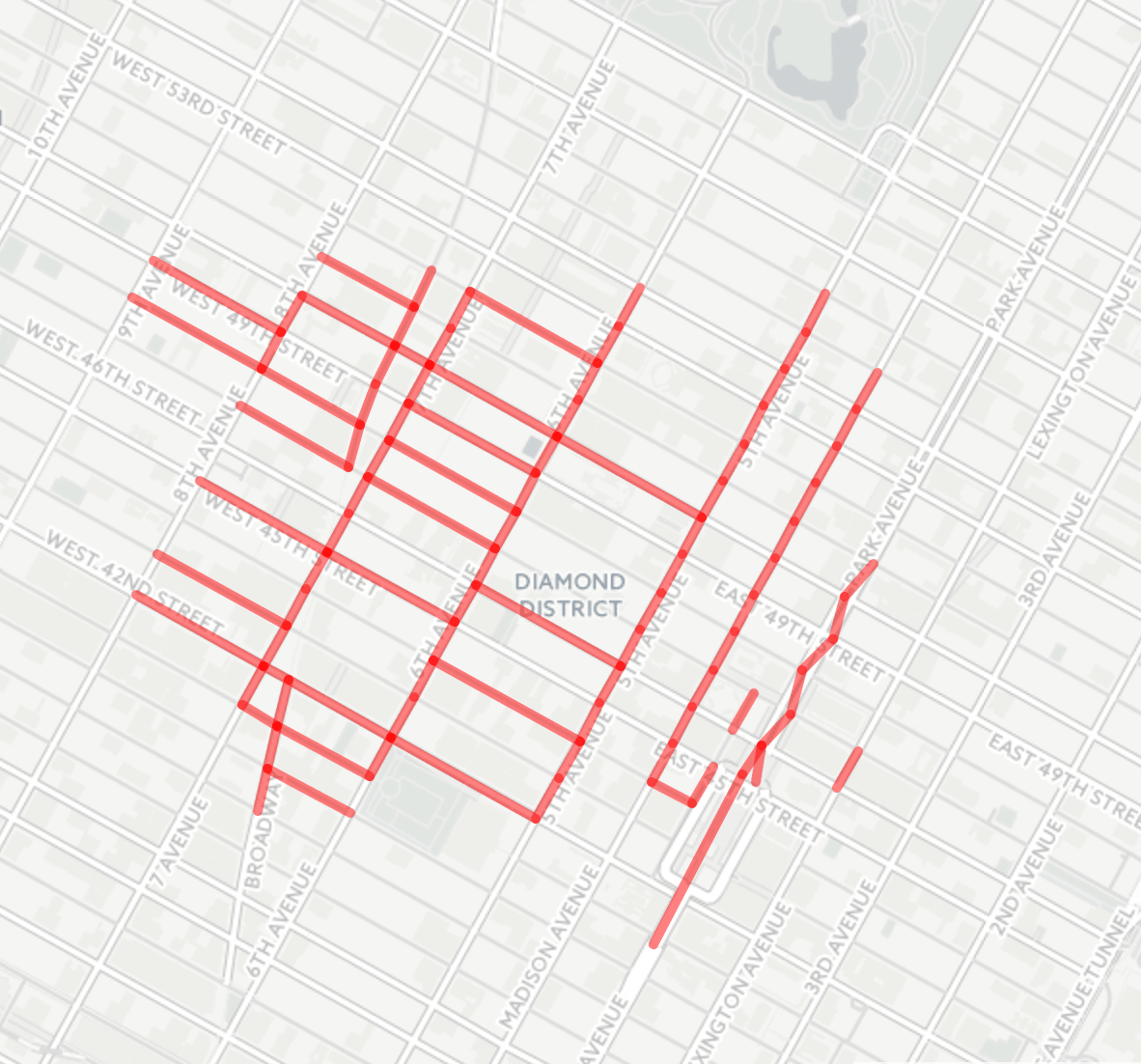}
\caption{Roads with speeds in $[2.5,\infty)$.  There are 5 connected 
components.}\label{fig:G_2p5}
\end{subfigure}
\begin{subfigure}{0.4\columnwidth}
\includegraphics[width=\columnwidth]{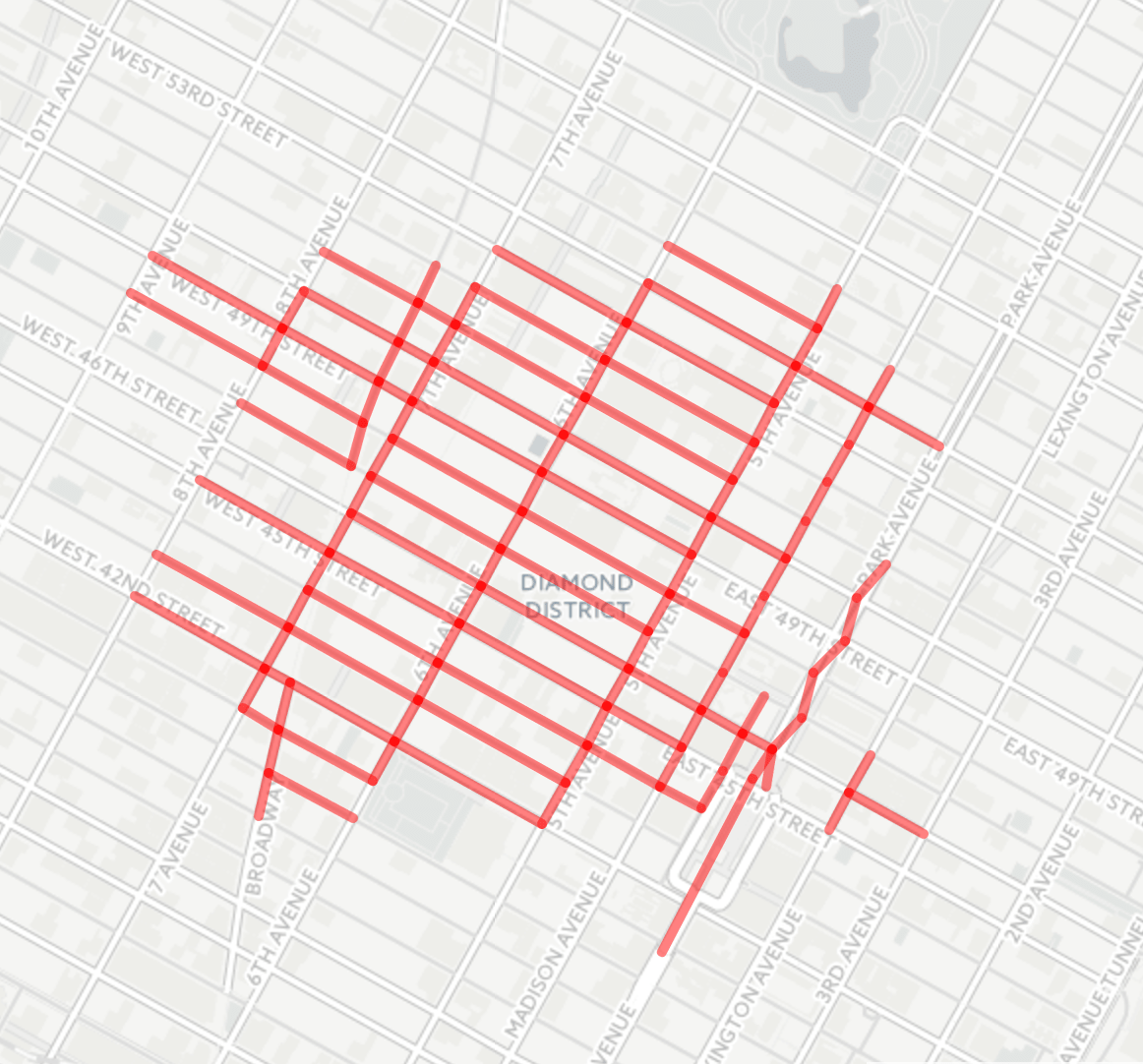}
\caption{Roads with speeds in $[2,\infty)$.  There are 2 connected 
components.}\label{fig:G_2}
\end{subfigure}
\begin{subfigure}{0.4\columnwidth}
\includegraphics[width=\columnwidth]{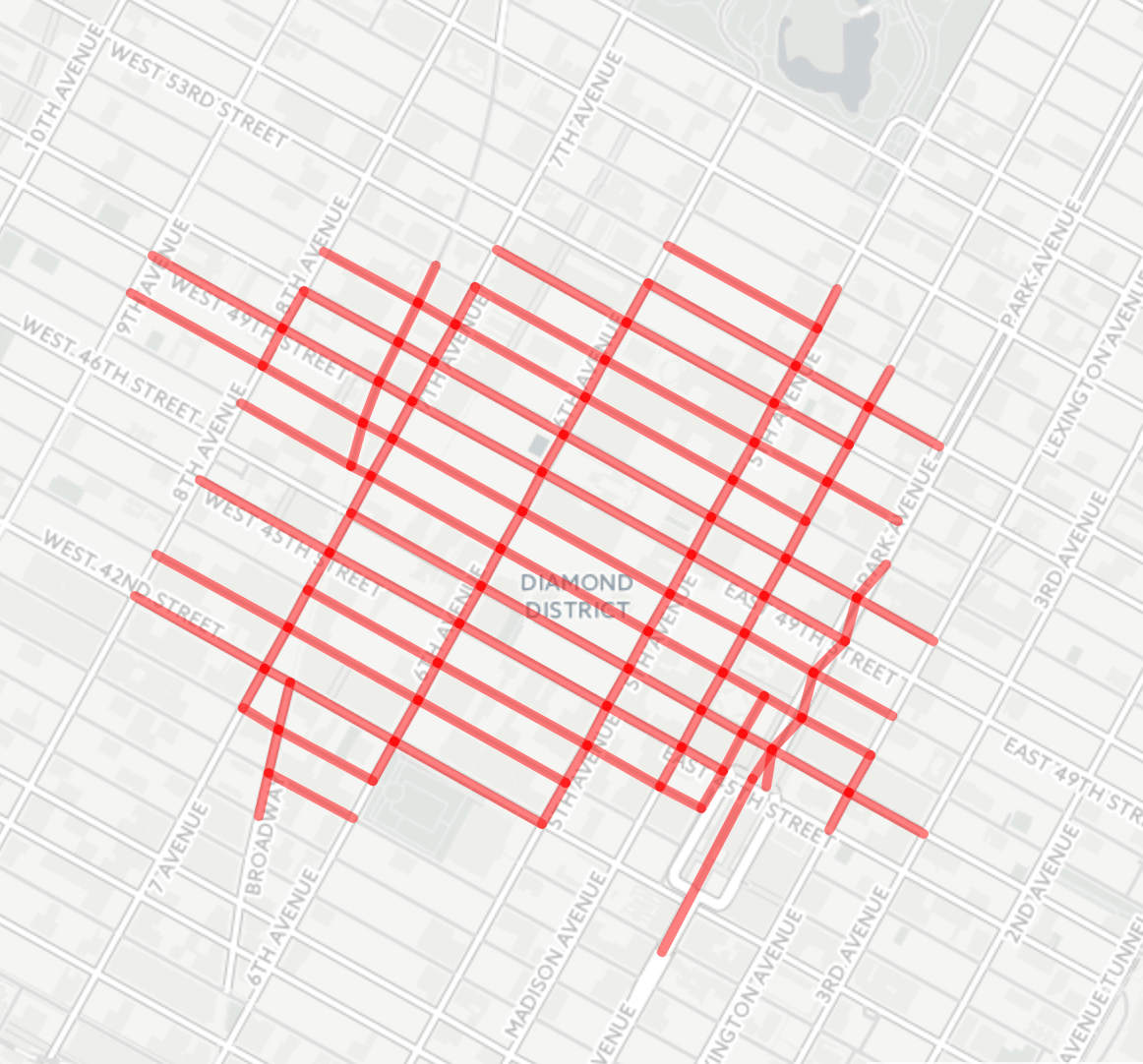}
\caption{Roads with speeds in $(0,\infty)$.  There is one connected 
component.}\label{fig:G_0}
\end{subfigure}
\caption{$G(\lambda)$ for $\lambda\in \{4,3.5,3,2.5,2,0\}$}
\end{figure}

We can reinterpret Figures \ref{fig:new_4toinf}--\ref{fig:new_0to2} via 
barcodes.  
Figure \ref{fig:G_4} consists of the roads with speeds faster than 4. 
We can add together the links of Figures 
\ref{fig:new_4toinf} and \ref{fig:new_3p5to4} and get the roads with speeds 
larger than 3.5; see Figure \ref{fig:G_3p5}.  
At the other extreme, if we augment Figure \ref{fig:new_2p5to3} with 
\emph{all} the roads with speeds at least 2.5, we get Figure \ref{fig:G_2p5}. 
Figure \ref{fig:DD} implicitly consists of all roads with nonnegative speeds.

\begin{figure}
\centering
\includegraphics[width=0.48\textwidth]{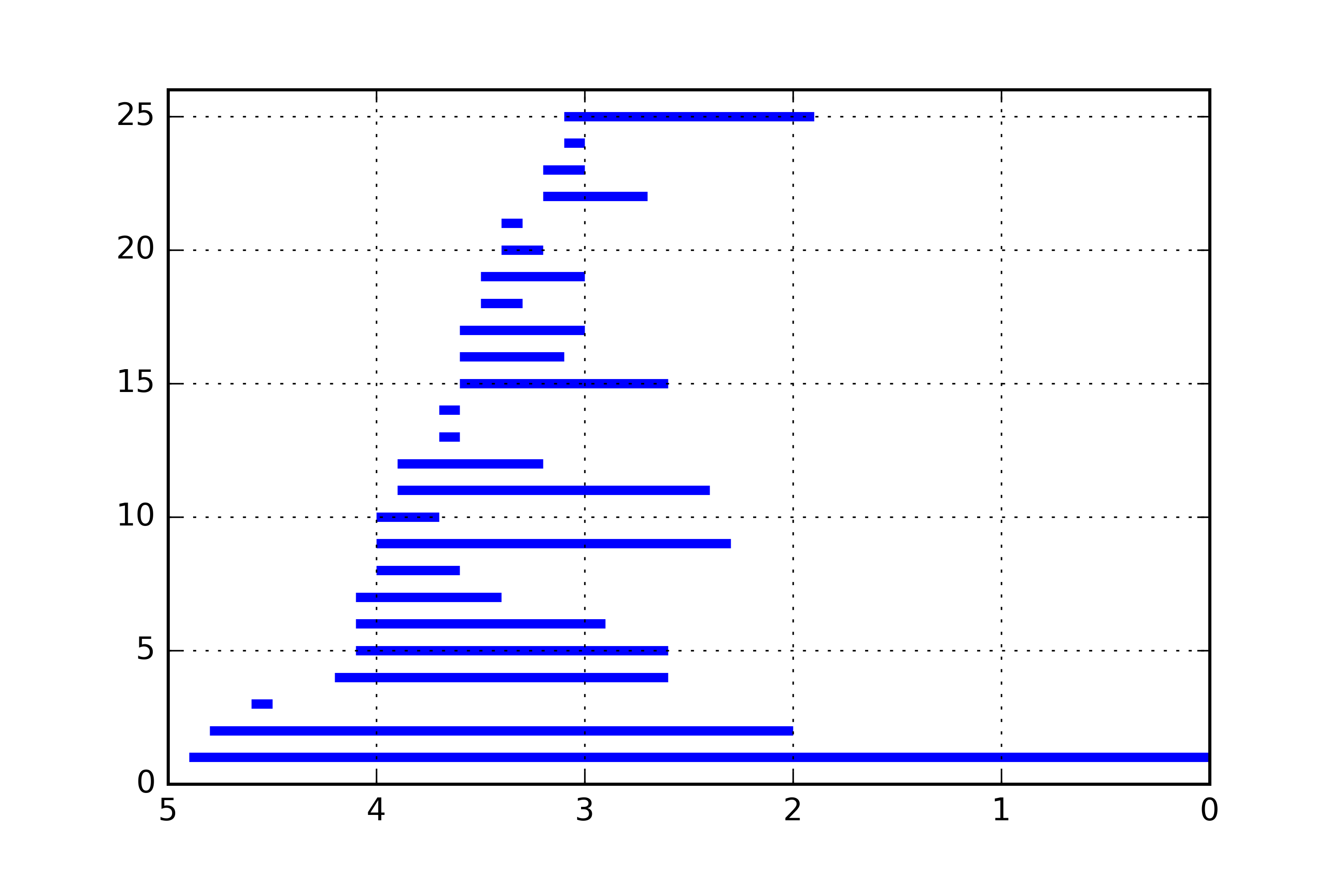}
\caption{barcode diagram for Diamond District.  Horizontal axis is $\lambda$, 
and vertical axis is the list index in barcode.}\label{fig:barcode}
\end{figure}

The barcode for the Diamond district is in Figure \ref{fig:barcode}.  
Reflecting Figures \ref{fig:G_4}--\ref{fig:G_0}, there are nine bars above 
$\lambda=4$, 14 bars above $\lambda=3.5$ (some start or end close to 3.5), 
10 bars above $\lambda=3$ (some end near 3), five bars above 2.5, 
two bars above 2, and one bar above 0.
There is a fair amount of instability near 3.5.  
Connected components appear, before being merged into other connected 
components.  
There are about 7 'long' bars (longer than length about 1).  
We can think of these as robust components.  
Three connected regions with speeds larger than about 2, but beyond, all of
the diamond district becomes one connected component.

\section{Conclusions and future work}
We have developed a barcode representation of congestion of
a reduced dataset. 
This visualization gives an alternate way
of looking at connectedness of congestion and gives us an understanding of 
robustness of congestion.

Applying this technique to Diamond District in New York shows that there are 
only a few small pockets where speeds are above 4 m/s, whereas in most of the 
region the speeds are between 2 and 4 m/s (Table~\ref{tab:summary}). 
It also 
shows that speeds are faster on North-South “Avenues” than on East-West 
“Streets”. 
These observations can be compared with neighboring areas and can 
show whether Diamond District is more or less congested than surrounding areas 
during observed times and if it offers greater uniformity in travel speeds.

This technique can also be applied to a much larger network of roads and in 
combination with other factors such as, time of day. 
For example, if a trip originates north of the Diamond District and concludes 
south of it during the evening rush hour, and the traveler has the option to 
either drive through or around the district, an understanding of congestion and 
variability of speeds during that time would assist in deciding an appropriate 
route. 
Overall, such analysis could also help identify areas in a city where traffic 
is significantly faster and areas that are severely congested, in turn assisting 
drivers avoid some areas, if possible, or plan for extra time. 
For planners, understanding the pattern, severity, and frequency of such 
congestion may help identify locations 
where infrastructure upgrade or congestion relief measures may be needed. 
For first responders, areas of chronic congestion may require additional 
resources or contingency plans. 

\begin{table}
\centering
\includegraphics[width=0.45\textwidth]{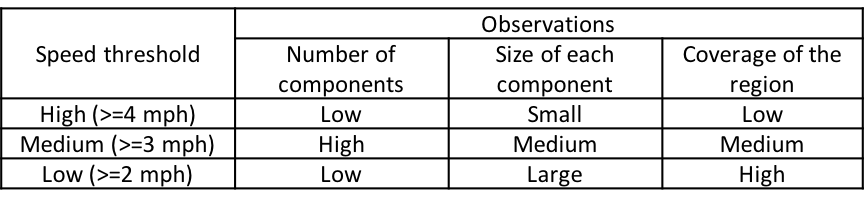}
\caption{Observations under different speed thresholds}\label{tab:summary}
\end{table}

Although our dataset is small enough that direct visual inspection of speeds on 
maps is feasible, our techniques
can give quantifiable information for larger and more complex datasets.

The notion of persistence can be taken in several new directions.  In this 
dataset, we have ignored directionality of streets (which in our case affects 
only a small number of streets).  
A proper treatment of direction requires deeper
theoretical innovations and will be developed elsewhere.

An interesting aspect of persistent homology lies in understanding the effects 
of boundaries.  
The roads in a network surround land which has a variety of uses.  
One might, for example, fill in land which has a certain population density.  
That would leadto thinking of roads as networks which allow inhabitants 
access to neighborhoods.  
One might also turn around and think of congestion as obstacles, much as in 
the robotics path planning work of \cite{bhattacharya2015persistent}. 
\balance

\bibliographystyle{unsrt}
\bibliography{references}

\end{document}